\documentclass[preprint, 12pt]{article}

\usepackage{geometry}
\usepackage{multirow}

\usepackage{csquotes}

\usepackage{ragged2e}
\usepackage{amsmath}

\usepackage{amsfonts,amssymb,amsthm,epsfig,subfigure,url,color}
\usepackage[mathscr]{euscript}

\newcommand{\nsum}[1][1.44]{\mathop{\vcenter{\hbox{
   \scalebox{#1}{$\displaystyle\sum$}}}}}

\usepackage{graphicx}
\graphicspath{ {./Figures/} }

\usepackage{caption}
\usepackage{titling}
\predate{ }
\postdate{ }
\date{ }

\providecommand{\keywords}[1]
{\begin{justify}
  \textit{Keywords:} #1
   \end{justify}}

\usepackage{sectsty}
\sectionfont{\fontsize{12}{15}\selectfont}

\subsectionfont{\fontsize{12}{15}\selectfont}

\subsubsectionfont{\fontsize{12}{15}\selectfont}

\usepackage{natbib}
\usepackage{multirow}
\usepackage{array}

\usepackage{float}
\floatstyle{plaintop}
\restylefloat{table}

\title{\textbf {The generalised Mooney space \\ for modelling the response \\ of rubber-like materials$^\dagger$}\vspace{-0.2cm}}

\author {\fontsize{13.5}{12}\selectfont Afshin Anssari-Benam\thanks{Corresponding author. Email address: afshin.anssari-benam@port.ac.uk (Afshin Anssari-Benam).} $^,$$^a$, Andrea Bucchi$^a$, \\
Michel Destrade$^b$, Giuseppe Saccomandi$^b$$^,$$^c$}

\begin{document}

\maketitle\vspace{-0.4cm}

\noindent\footnotesize $^a$ Cardiovascular Engineering Research Lab (CERL), School of Mechanical and Design Engineering, University of Portsmouth, Anglesea Road, Portsmouth PO1 3DJ, United Kingdom. \\ 

\noindent\footnotesize $^b$ School of Mathematics, Statistics and Applied Mathematics, NUI Galway, University Road, Galway, Ireland. \\

\noindent\footnotesize $^c$ Dipartimento di Ingegneria, Università degli studi di Perugia, Via G. Duranti, Perugia 06125, Italy. \\
\
\begin{center}
                       \noindent\normalsize $^\dagger$  Dedicated to Millard Beatty on the occasion of his 90$^{\textnormal{th}}$ birthday.\\ 
\end{center}

\bigskip

\hrule

\medskip

\normalsize 

\renewenvironment{abstract} 
{\begin{justify}\textbf{Abstract\\}\ignorespaces}
{\end{justify}}

\begin {abstract}

Soft materials such as rubbers, silicones, gels and biological tissues have a nonlinear response to large deformations, a phenomenon which in principle can be captured by hyperelastic models. 
The suitability of a candidate hyperelastic strain energy function is then determined by comparing its predicted response to the data gleaned from tests and adjusting the material parameters to get a good fit, an exercise which can be deceptive because of nonlinearity. 
Here we propose to generalise the approach of Rivlin \& Saunders [Phil Trans A 243 (1951) 251-288] who, instead of reporting the data as stress against stretch, manipulated these measures to create the `Mooney plot', where the Mooney-Rivlin model is expected to produce a linear fit. 
We show that extending this idea to other models and modes of deformation (tension, shear, torsion, etc.) is advantageous, not only (a) for the fitting procedure, but also to (b) delineate trends in the deformation which are not obvious from the raw data (and may be interpreted in terms of micro-, meso-, and macro-structures) and (c) obtain a bounded \textit{condition number} $\kappa$ over the whole range of deformation; a robustness which is lacking in other plots and spaces.

\end{abstract}


\keywords{soft materials; nonlinear elasticity; curve fitting; Mooney plot; condition number}


\medskip

\hrule

\vspace{0.2cm}

\section{Introduction}
\label{section1}

\vspace{-0.4cm}

\

A great variety of models are available in the literature to describe the nonlinear behaviour of rubber-like materials in large deformations. 
Once an \textit{appropriate} strain energy function $W$ is selected for a given specimen, the important next step is to fit the model's predictions to the acquired experimental data. 
In a good case scenario, the outcomes of an \textit{accurate} fitting produce low relative errors (e.g., \citeauthor{MethodicalFit}, \citeyear{MethodicalFit}), ideally provide a single set of model parameter values that can capture various deformation modes of the specimen  (e.g.,  \cite{ABBModel}, \citeauthor{RCT}, \citeyear{RCT}), and optimally result in a model that remains stable beyond the collected range of experimental data (e.g., \citeauthor{Yeoh97} \citeyear{Yeoh97}). 
Therefore, in view of the \enquote{\textit{Hauptproblem}} of nonlinear elasticity \citep{Truesdell56}, the fitting process is of utmost importance.  

To address these issues, \cite{OgdenFitting} systematically studied the fitting of hyperelastic models to experimental data, with a particular focus on the problem of the uniqueness, or lack thereof, of the optimal fit. 
In the wake of that seminal paper, studies such as those by \cite{ABBModel} or \cite{RCT} sought to obtain the optimal result by fitting the models to various deformation datasets of each specimen simultaneously, as opposed to using data from a single deformation mode only, which is commonly practiced in the literature. 
In the same spirit, \cite{Yanfoam} proposed that, when modelling the simple shear of elastomeric foams, the stress components along the inclined surface should also be considered, and that the model should simultaneously be fitted to all those stress components. 
\color{black}Appendix B of \cite{I2Paper} also provides further comparisons between various fitting strategies. 

As reviewed recently by \cite{MethodicalFit}, however, an illuminating alternative approach to the identification of the model parameters is the use of the so-called \textit{Mooney space} for data fitting. This approach was first introduced by \cite{RS} for application to the data obtained from uniaxial tension tests on rubbers in conjunction with the Mooney-Rivlin model. 
Instead of using the traditional Cauchy or engineering spaces (where the stress is reported against the stretch), \cite{RS} transformed their data into the Mooney space, where the Mooney-Rivlin model predicts that the data points should be along a straight line. 
Modelling the uniaxial data in the Mooney space highlights some nuanced aspects of the deformation and modelling trends which are hidden in the Cauchy or engineering spaces. These include the magnification of the model performance in small to medium deformation ranges, as noted by \cite{PucciandSacco}, or the identification of different deformation regimes which are each associated with different mesoscopic phenomena and can be clearly delineated in the Mooney space as demonstared by \cite{MethodicalFit}. 
Importantly, the transformation of the uniaxial data and model formulation into the Mooney space allowed \cite{RS} to reduce the fitting procedure to a \textit{linear regression} problem. 
This advantageous aspect can be demonstrated as follows.

The well-known representation formula of the Cauchy stress for incompressible isotropic materials reads:  
\begin{equation} \label{eq1}
\textbf{T}=-p\textbf{I}+2W_1\textbf{B}-2W_2\textbf{B}^{-1},
\end{equation}
where $p$ is the arbitrary Lagrange multiplier enforcing the condition of incompressibility, $\textbf{I}$ is the identity tensor, $\textbf{B}$ is the left Cauchy-Green deformation tensor, and $W_1$ and $W_2$ are the partial derivatives of the strain energy function $W$ with respect to $I_1=\textnormal{tr}\textbf{B}$ and $I_2=\textnormal{tr}\textbf{B}^{-1}$, the first and second principal invariants of $\textbf{B}$, respectively, with $I_3=1$ due to incompressibility. 
For uniaxial deformations, $\textbf{B}= \textnormal{diag}\left(\lambda^2,\lambda^{-1},\lambda^{-1}\right)$, where $\lambda$ is the uniaxial stretch.
By setting $T_{22}=T_{33}=0$ as the boundary conditions to establish $p$, the relationship between the uniaxial Cauchy stress $T_{11}$  and the stretch $\lambda$ is obtained as:
\begin{equation} \label{eq3}
T_{11}=2\left(\lambda^2-\cfrac{1}{\lambda}\right) \left(W_1+\cfrac{1}{\lambda}W_2\right).
\end{equation}
Equation (\ref{eq3}) is then re-written in terms of the engineering stress $P = T_{11}\lambda^{-1}$ as: 
\begin{equation} \label{eq4}
P=2\left(\lambda-\cfrac{1}{\lambda^2}\right) \left(W_1+\cfrac{1}{\lambda}W_2\right).
\end{equation}
Instead of reporting the Cauchy plot ($T_{11}$ against $\lambda$) or the engineering plot ($P$ against $\lambda$), Rivlin and Saunders created the \textit{Mooney plot} by transforming Equation (\ref{eq4}) into:
\begin{equation} \label{eq5}
\mathscr{M}=W_1+\zeta W_2\thinspace,
\quad \text{where} \quad   \zeta=\cfrac{1}{\lambda}, \qquad
  \mathscr{M}=\cfrac{P}{2\left(\lambda-\cfrac{1}{\lambda^2}\right)},    
\end{equation}
and reported the data as $\mathscr{M}$ against $\zeta$. Then, on using the Mooney-Rivlin model,  
\begin{equation} \label{MR}
W_\text{MR} = \tfrac{1}{2} C_1 (I_1-3) + \tfrac{1}{2} C_2 (I_2-3),
\end{equation}
where $C_1$ and $C_2$ are constants, it becomes clear from Equation \eqref{eq5} that: 
\begin{equation} \label{eq7}
\mathscr{M}_\text{MR}=C_1+ C_2 \zeta,
\end{equation}
so that fitting the Mooney-Rivlin model to uniaxial data becomes a matter of straightforward \textit{linear regression} in the Mooney space; a clear advantage given the computational power available at the time.

Interestingly, \cite{RS} did not provide an explicit rationale for presenting the uniaxial deformation data in the Mooney space.
Whatever the initial motivation, however, the concept of the Mooney plot has since been used frequently as an analytical tool to study the mechanical behaviour of rubbers. 
Some notable examples of the direct use of the Mooney plot include the work of \cite{GentT} to develop their logarithmic $I_2$ model, \cite{FukahoriSeki} to experimentally evaluate the values of $W_1$ and $W_2$, and \cite{McKenna1} to analyse and model the mechanical behaviour of swollen rubbers.

However, in using the \textit{classical} Mooney plot of Equation (\ref{eq5}), due care must be exercised for strain energy functions $W$ with functional forms other than that of the Mooney-Rivlin model. 
In general, there is no guarantee that the functional form of $\mathscr{M}$ for different models \color{black}in the classical Mooney space will be conducive to a standard linear regression analysis. 
Indeed, it can be demonstrated that for some $W$ functions, a direct linear regression link in the Mooney space cannot be established. 
Although they did not explicitly expand on this point, it appears that \cite{GentT} were the first to recognise this fact by adopting a new domain for $\zeta$ and $\mathscr{M}$ instead of that in Equation (\ref{eq5}) originally used by \cite{RS}; see their figure 1. 
The same trait may also be found in the later works of McKenna and co-workers, who defined various Mooney domains for different models of swollen rubbers; see, e.g., \cite{McKenna1} and \cite{McKenna2}. 
Hence, it is clear that the classical Mooney space does not provide the generality of framework for the adoption and application of many of the existing models in the literature. 
In addition, the domain of application of the Mooney space has thus far been limited to the uniaxial deformation.

Accordingly, in this paper we present a systematic approach to define a \textit{generalised Mooney space} in which a linear regression is achieved for various strain energy functions $W$ by devising appropriate corresponding measures of $\mathscr{M}$ and $\zeta$. 
It is therefore our aim to recast the classical Mooney plot into a {canonical} form, namely the \textit{generalised Mooney plot}, which allows the transformation of the fitting process into a standard linear regression problem for several strain energy functions $W$.  
We demonstrate that for some $W$ functions the classical Mooney plot is not always the ideal means to represent the modelling results versus the experimental data. 
Depending on the particular form of $W$, we show in \S2 that alternative functional forms of $\mathscr{M}$ may be formulated within the \textit{generalised Mooney space} which result in a proper linear regression for the demonstration of the model fittings to the data. 
We use the simple extension deformation as our point of departure. 
Then in \S3 we extend the idea of the generalised Mooney spaces and plots to other standard deformation modes (e.g., equi-biaxial tension, pure shear, simple shear and simple torsion).
Finally, we provide concluding remarks in \S4.


\section{Generalised Mooney spaces and generalised Mooney plots}
\label{section2}


To showcase the mathematical concept of the \textit {generalised Mooney space}, or GMS for short, here we take the archetypal example of simple tension deformation as our starting point. 
The kinematics of this deformation is described by the deformation gradient $\textbf{F}=\text{diag}(\lambda, 1/\sqrt{\lambda},1/\sqrt{\lambda})$, and thus by the single variable $\lambda$, the principal stretch in the direction of tension. 
Then, in uniaxial deformation tests, ordered pairs $\{\lambda_i,P_i\}$ are measured and collected, where $P$ is the engineering stress. 

We define the GMS as a space where, for a given model with material parameters $C_j$,  two data sets  $\mathscr{M}$ and $\zeta$ can be constructed from the ordered pairs $\{\lambda_i,P_i\}$ to yield a relationship in the form:
\begin{equation} \label{eq8}
\mathscr{M}=  \sum_{j  \in A} C_j \zeta^{j},
\end{equation}
(where $A$ is a finite subset of $\mathbb{Q}$), so that the $C_j$'s are determined by a linear curve-fitting exercise. 

We  show below that this representation is highly illuminating from the perspective of highlighting deformation regimes and trends that are not apparent in other spaces. 
More crucially, however, it can be demonstrated using the \emph{condition numbers} associated with the stress quantities $P$, $T$, $\mathscr{M}$ that curve fitting in the GMS is more advantageous than in the classical Cauchy or engineering spaces.  
Recall that the condition number $\kappa$ of a one-variable function $f=f(x)$ is $\kappa=|x f'/f|$.
The condition number is in effect a measure of how sensitive a regression is to perturbations in the data points \citep{Belsley1980}. 
In the context of applying mathematical models to the experimental data, the condition number $\kappa$ may be interpreted as how robust  the obtained fit is, given the degree of experimental errors and uncertainties inherently present in the deformation datasets of soft solids. 
By definition, the lower the value of $\kappa$, the less sensitive the regression is to perturbations and hence the more robust the fitting result is. 
It can be shown on a case-by-case basis that the $\kappa$ number obtained by fitting a model to the experimental data in the GMS is \textit{a priori} lower than that obtained using the same model in either Cauchy or engineering spaces.

Let us take the Mooney-Rivlin model as an example. 
We find from Equations \eqref{eq3} and \eqref{eq4} that: 
\begin{equation} \label{eq110cmp}
\kappa_T = \cfrac{C_1\lambda(2\lambda^{3} + 1) + C_2(\lambda^{3}+2)}{(\lambda^{3}-1)(C_1\lambda+C_2)}\quad, \qquad
\kappa_P = \cfrac{C_1\lambda(\lambda^{3}+2)+3C_2}{(\lambda^{3}-1)(C_1\lambda+C_2)},
\end{equation}
showing that in the Cauchy and engineering spaces the data is presented in such a way that the condition number cannot be smaller than 1 $\left(\lim\limits_{\lambda \to \infty} \kappa_T, \kappa_P = 1\thinspace\right)$ and becomes extremely large at small stretches $\left(\lim\limits_{\lambda \to 1} \kappa_T, \kappa_P = \infty\thinspace\right)$. 
In contrast, in the GMS the condition number $\kappa$ is found from Equation \eqref{eq7} as: 
\begin{equation} \label{eq112}
\kappa_\mathscr{M} = \cfrac{C_2 \zeta}{C_1 + C_2\zeta}\thinspace, \quad \text{ so that } \quad 0<\kappa_\mathscr{M}\leq1.
\end{equation}
Hence the condition number $\kappa$ obtained by fitting the Mooney-Rivlin model to the experimental data  in the GMS is never greater than 1 (for small stretches, as $\zeta \to 1$). Therefore, the GMS facilitates obtaining more robust fits to the data compared with the Cauchy or engineering spaces.

We now present examples of the transformation of various models and the data into the GMS and obtain the ensuing \textit{generalised Mooney plots} (GMPs) in this space, contrasting them with the plots in the classical Mooney space.
\color{black}Note that the term `model' in this work is used exclusively in reference to the strain energy function $W$. The modelling tool is the GMS defined by $\mathscr{M}$ and $\zeta$ via Equation (\ref{eq8}).
\color{black}The list of model examples in this section is of course not exhaustive. 
The $\mathscr{M}$ function defined in Equation (\ref{eq8}) may be tailored  to a variety of  strain energy functions $W$, where the concept of GMS is well defined. 
However, we also note that there exist strain energy functions $W$ that are not amenable to a linear regression transformation within the GMS framework. 
Such strain energy functions include, for example, the \cite{ogden1972large} model:
\begin{equation}
W_\text{O} = \sum_{i=1}^N \frac{\mu_i}{\alpha_i} (\lambda_1^{\alpha_i} + \lambda_2^{\alpha_i} + \lambda_3^{\alpha_i} -3),
\end{equation}
(where the $\lambda$s are the principal stretches),
the model proposed recently by \cite{ABBModel}:
\begin{equation} \label{eq27}
W_\text{ABB}=\mu N\left[\cfrac{1}{6N}(I_1-3)-\text{ln}\left(\cfrac{I_1-3N}{3-3N}\right)\right],
\end{equation}
or the Gent+Gent model proposed by \cite{PucciandSacco}:
\begin{equation} \label{eq35}
W_\text{GG}=-\cfrac{\mu J_m}{2}\medspace\textnormal{ln}\left(1-\cfrac{I_1-3}{J_m}\right)+C_2\text{ln}\left(\cfrac{I_2}{3}\right),
\end{equation}
where $\mu_i$, $\alpha_i$, $\mu$, $N$, $C_2$ and $J_m$ are material parameters.

The GMPs presented henceforth have been obtained by minimising the \emph {relative} error. We emphasise that it is crucial in curve fitting exercises to minimise the relative error, and not the absolute error (as is often done by default in the literature).
The absolute error changes from one stress measure to another, and thus a curve fitting exercise based on absolute errors would predict different optimal material constants for the same test depending on whether the experimentalist chose to report, for example, the Cauchy or the engineering stresses. 
Accordingly, here we conduct the linear optimisation procedure based on minimising the relative error, because it yields the same optimal set of material constants independently of the stress measure \citep{MethodicalFit}. 
This is easily achieved by any software optimisation code, simply by using a classical weighted Least-Square procedure to minimise the relative residual sum defined here as: $\nsum[1.2]_i \left[(\mathscr{M}^\text{model}_i - \mathscr{M}^\text{experiment}_i)/\mathscr{M}^\text{experiment}_i\right]^2$.


\subsection{Yeoh and polynomial neo-Hookean models}


Consider first the cubic \emph{Yeoh model} \citep{Yeoh1990}, often used to model the deformation of rubbers:
\begin{equation} \label{eq9}
W_\text{Y} = C_1\left(I_1-3\right)+C_2\left(I_1-3\right)^2+C_3\left(I_1-3\right)^3.
\end{equation}
From Equation (\ref{eq4}) it follows that:
\begin{equation} \label{eq10}
P=2(\lambda- \lambda^{-2}) \left[C_1+2C_2\left(I_1-3\right)+3C_3\left(I_1-3\right)^2\right]\thinspace.
\end{equation}
Because in uniaxial tension $I_1=\lambda^2 + 2\lambda^{-1}$, the function $\mathscr{M}$ for the Yeoh model is thus:
\begin{equation} \label{eq11}
\mathscr{M}_\text{Y}=C_1+2C_2\zeta+3C_3\zeta^2,
\quad 
\text{where} \quad
\zeta := \lambda^2+ 2\lambda^{-1}-3.
\end{equation}
A generalisation of this model is the \emph{polynomial neo-Hookean model} (pnH) as:
\begin{equation} \label{eq13}
W_\text{pnH}(I_1)=\sum^{n}_{i=1} C_i (I_1-3)^i.
\end{equation}
This model recovers the Yeoh model and also the Taylor series approximation of the many existing generalised neo-Hookean strain energy functions in the literature; see, e.g., \cite{BoyceComparison} for a review. 
In a similar manner as to the Yeoh model here we find that:
\begin{equation} \label{eq14}
P=2(\lambda - \lambda^{-2} )\sum^{n}_{i=1} i\thinspace C_i (\lambda^2 + 2 \lambda^{-1} - 3)^{i-1},
\end{equation}
which leads to the following $\mathscr{M}$ function in the GMS:
\begin{equation} \label{eq15}
\mathscr{M}_\text{pnH}=\sum^{n}_{i=1} i\thinspace C_i \thinspace \zeta^{i-1}, \quad 
\text{where} \quad
\zeta:=\lambda^2 + 2 \lambda^{-1}-3.
\end{equation}

\begin{figure}[h]
    \advance\leftskip-2cm
             \includegraphics[width=19cm, height=6.75cm]{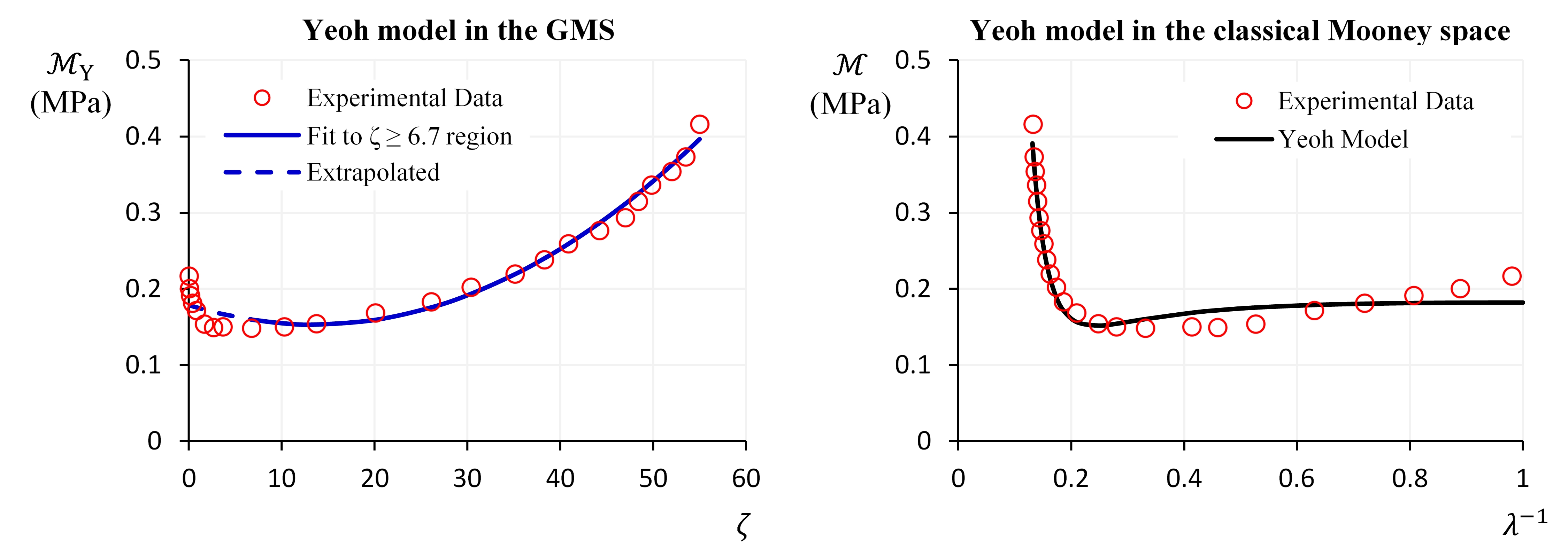} 
     \caption{The Yeoh model in the GMS (left panel) and the classical Mooney space (right panel). 
The former provides a better resolution for inspecting the suitability of the model in describing the data than the latter. 
The results in the GMS indicate that the Yeoh model is not suitable for capturing the deformation at all stretches, because the data there cannot be captured by a quadratic variation as expected. 
The model is suitable to describe the deformation at  higher stretches, say $\zeta \geq 6.7$ ($\lambda \geq 3$). 
Experimental data is from the uniaxial deformation due to \cite{TreloarData}.}
\label{Fig1}
\end{figure}

On using the uniaxial data due to \cite{TreloarData}, Figure \ref{Fig1} presents the data plots in both the GMS and classical Mooney space for the Yeoh model. 
Compared with the classical Mooney space, it is observed that the GMS provides a more versatile tool to inspect the suitability of the model in describing the experimental data. 
The trend of the data points in this space is first a decrease in $\mathscr{M}_\text{Y}$, then  a minimum and finally an increase with $\zeta$. 
The transformation into the GMS therefore makes it clear that the parabolic form of the Yeoh model in Equation \eqref{eq11} cannot account for the asymmetric distribution of data near $\zeta=0$, i.e., at small deformations. 
However, this parabolic shape is clearly more effective for larger $\zeta$, say $\zeta \geq 6.7$, which is equivalent to $\lambda \geq  3$. 
This degree of resolution for delineating various trends in the dataset and the modelling results is not provided by the classical Mooney space, see Figure \ref{Fig1}(b).  

\color{black}


\subsection{Gent-Thomas model}


Now we consider the classic \emph{Gent-Thomas model} \citep{GentT}: 
\begin{equation} \label{eq23}
W_\text{GT}=C_1\left(I_1-3\right)+C_2\thinspace\text{ln}\left(\cfrac{I_2}{3}\right).
\end{equation}
For this model we have:
\begin{equation} \label{eq24}
P=2\left(\lambda-\cfrac{1}{\lambda^2}\right)\left(C_1+\cfrac{C_2}{2\lambda^2+\lambda^{-1}}\right),
\end{equation}
noting that in uniaxial deformation $I_2=\lambda^{-2}+2\lambda$. 
It is straightforward to see that we can form a linear regression problem as:
\begin{equation} \label{eq25}
\mathscr{M}_\text{GT} = C_1 +  C_2 \zeta\thinspace, \quad \text{where} \quad \zeta:=\cfrac{1}{2\lambda^2+\lambda^{-1}}\thinspace.
\end{equation}

\begin{figure}[H]
    \centering
             \includegraphics[width=0.7\textwidth]{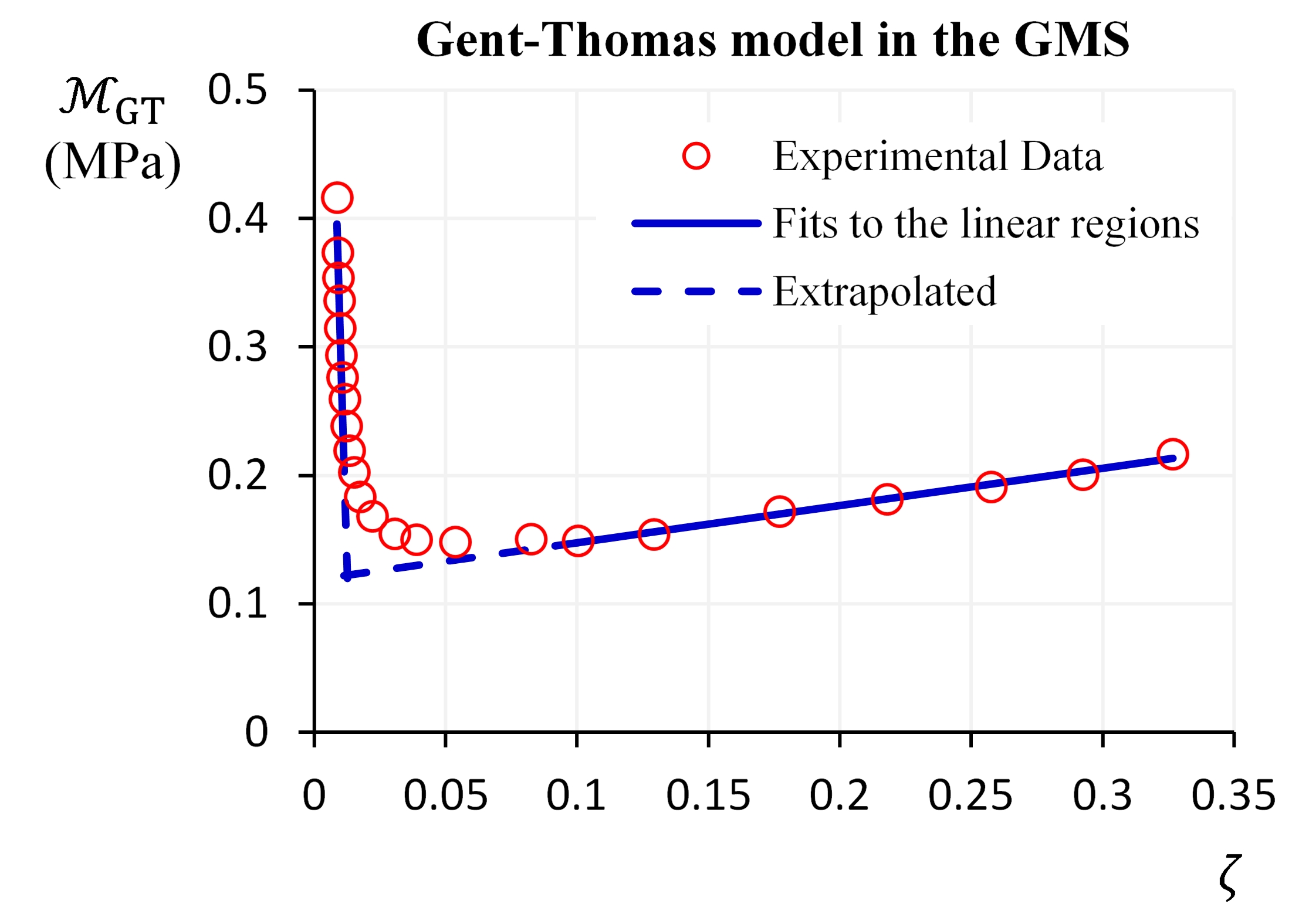} 
     \caption{The Gent-Thomas model in the GMS, where it is expected to give a linear equation. 
Clearly, here, there is no single line trend that can cover the entire range of stretches in the experimental data. 
However, with the help of the GMS it is possible to identify the linear region(s) within the data and direct the focus of the modelling campaign only on these regions when using the Gent-Thomas model. 
The lines represent the fitting results for the two identifiable linear regions. Experimental data is from the uniaxial deformation due to \cite{TreloarData}.}
\label{Fig2}
\end{figure}

Figure \ref{Fig2} illustrates the transformation of this model in the GMS using Equation \eqref{eq25} and its correlation with the uniaxial experimental data of \cite{TreloarData}. 
For this model, $\mathscr{M}_\text{GT}$ in the GMS coincides with that of the classical Mooney space.
However, note that by definition, $\zeta$ varies in the range $0<\zeta<1$ in the classical Mooney space, whereas the domain of $\zeta$ defined in the GMS via Equation \eqref{eq25} for this dataset is $0<\zeta<1/3$. Therefore, compared with the classical Mooney space, the GMS provides a better magnification of the data trends and the performance of the model. 


It is evident from Equation \eqref{eq25} and the GMP in Figure \ref{Fig2} that the Gent-Thomas model is best suited for application to the linear regions of the data, and not to the entire stretch range. 
Using the GMS it is possible to identify two regions within which the data trend may be considered as linear: (i) a region where $\zeta$ varies between $0.1<\zeta<1/3$ (corresponding to $1<\lambda<2.18$); and (ii) a region where $\zeta<0.01$ (corresponding to $\lambda>6.85$).
By focusing the modelling effort in these two linear regions, the model is then capable of providing a suitable description of the data, as shown by the lines in Figure \ref{Fig2}.
However, we also note that the provided linear fit by the Gent-Thomas model for $0.01>\zeta$ is obtained only when $C_2<0$ (as the slope is clearly negative there), which may not be physically valid and may lead to the loss of ellipticity. 
Therefore, for a physically valid result, the GMS suggests that the application of the Gent-Thomas model should be limited to a certain range of deformation, in this case to when $0.1<\zeta<1/3$ (or $1<\lambda<2.18$). 
These trends are not easily distinguishable in the Cauchy or engineering spaces.


\subsection{Gent model} 

\begin{figure}[h]
    \advance\leftskip-2cm
             \includegraphics[width=19cm, height=6.85cm]{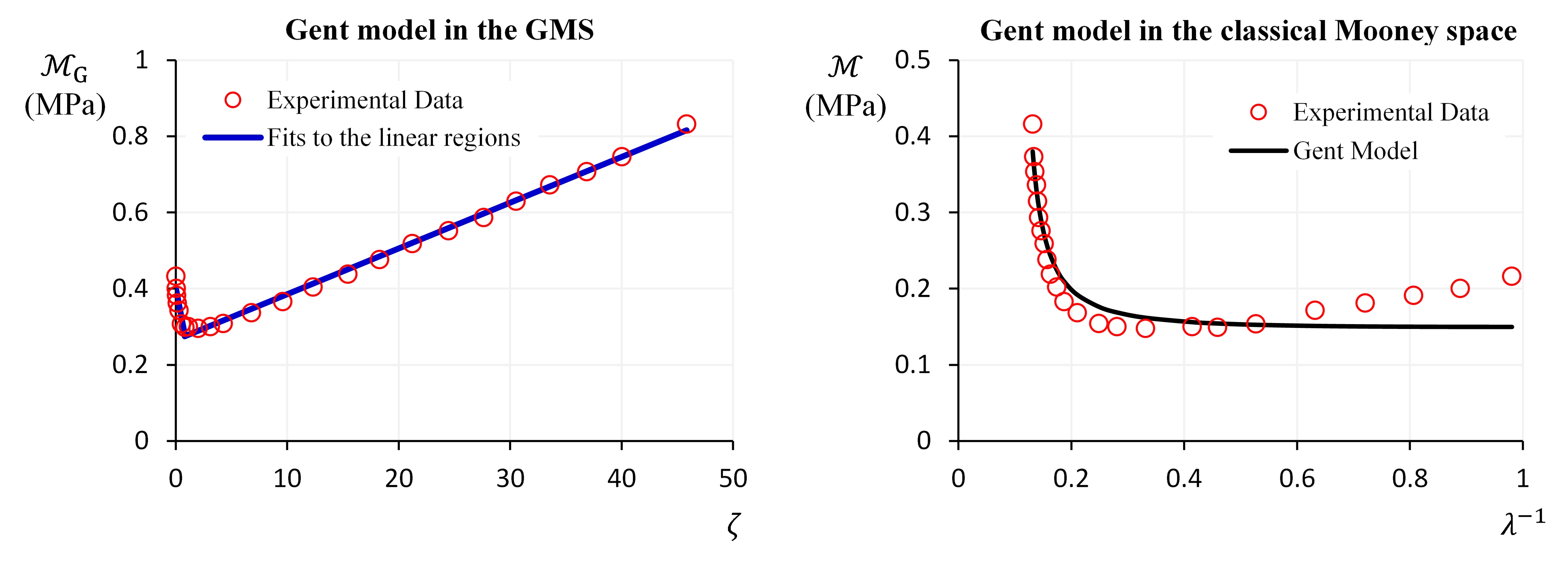} 
     \caption{The performance of the Gent model in the GMS (left panel) and the classical Mooney space (right panel). 
It is possible to identify a V-shape trend in the experimental data in the GMS. In either branches of this V-shaped pattern, then, the Gent model is able to provide a linear fit, but clearly not to the whole range of the data. The blue lines show the provided best lines of fit to the data in each branch of this V.
Experimental data is from the uniaxial deformation due to \cite{TreloarData}. 
}
     \label{Fig3}
\end{figure}

Next, we consider the \emph{Gent model} \citep{Gent1996}:
\begin{equation} \label{eq16}
W_G=-\cfrac{\mu J_m}{2}\medspace\textnormal{ln}\left(1-\cfrac{I_1-3}{J_m}\right),
\end{equation}
where $\mu$ is the infinitesimal shear modulus and $J_m$ is the stiffening parameter.
On using Equation (\ref{eq4}) we find:
\begin{equation} \label{eq17}
\cfrac{P}{\lambda - \lambda^{-2}} = \cfrac{\mu}{1 - (1/J_m)\left(\lambda^2+2\lambda^{-1}-3\right)}\thinspace.
\end{equation}
It then follows that: 
\begin{equation} \label{eq19}
\cfrac{P}{\lambda - \lambda^{-2}} -  \cfrac{P}{\lambda - \lambda^{-2}} \thinspace \cfrac{\lambda^2+2\lambda^{-1}-3}{J_m} = \mu\thinspace,
\end{equation}
so that we arrive at the linear regression problem: 
\begin{equation} \label{eq21}
\mathscr{M}_\text{G} = \mu + \cfrac{1}{J_m}\zeta\thinspace,
\end{equation}
where: 
\begin{equation} \label{eq22}
\mathscr{M}_\text{G}:=  \cfrac{P}{\lambda-\lambda^{-2}}\thinspace, \qquad \mathscr{\zeta}:= \cfrac{P\left(\lambda^2 + 2\lambda^{-1} -3\right)}{\lambda - \lambda^{-2}}\thinspace.
\end{equation}
Here, $\mathscr{M}_\text{G}$  and $\zeta$ are a mixed set of stretch and engineering stress measures. 
Clearly, in this case it is not straightforward to ascribe a direct physical interpretation to the GMS. 
However, the GMS is still useful for investigating the goodness of the fit provided by the Gent model. 
If the data is linear in $\zeta$ within this space, we find directly the value of $J_m$ as the inverse slope of the line and of $\mu$ as its intercept. 
Moreover, the fitting exercise has been transformed from a nonlinear procedure, see Equation \eqref{eq17}, to a much more straightforward linear one.
Figure \ref{Fig3} shows Treloar's data in the GMS associated with the Gent model.


Interestingly, the transformation of the data into the GMS reveals a V-shaped pattern in the experimental data. 
It is clear that the model cannot capture this V-shaped pattern, as Equation \eqref{eq21} is the equation of a single line. 
Instead, however, the GMS allows us to perform a good linear fit on either branches of the V-shaped data, depending on whether we wish to model the early or later regimes of extension. 
The apex of this V-shaped pattern is located at $\zeta \simeq 0.8$, corresponding to $\lambda \simeq 2.18$. 
In either sides of this apex, the data in the GMS indicate a linear trend, which then the Gent model in Equation \eqref{eq21} is able to provide a good fit to, see the lines in the GMP of Figure \ref{Fig3}(a). 
However, note that for $\zeta<0.8$, or equivalently $\lambda<2.18$, the slope of the fitted line is negative, which in turn indicates that the value of the parameter $J_m$ must be negative too. 
From a meso-structural perspective, a negative $J_m$ is not physically realistic, and thus the Gent model may not be suitably used in this context for modelling the deformation within the small stretch regime.  
These trends, and analyses, are not lucid in the classical Mooney space, as shown in Figure \ref{Fig3}(b), and will be even less so in the Cauchy and engineering spaces. 


\subsection{Fung model}


Another example of a widely used strain energy function $W$ in the literature, particularly pertaining to the biomechanics of soft tissues, is provided by the exponential \emph{Fung-Demiray model} \citep{beatty1987topics}:
\begin{equation} \label{Fung1}
W_\text{FD} = \frac{\mu}{2b}\thinspace\left[\text e^{b(I_1 - 3)}-1\right],
\end{equation}
where $\mu$ is the infinitesimal shear modulus and $b$ is the stiffening parameter.
On transforming this model into the GMS we also find a linear formula as:
\begin{equation} \label{Fung2}
\mathscr{M}_\text{FD} = \ln \mu + b\zeta\thinspace,  
\end{equation}
where:
\begin{equation}\label{Fung3} 
\mathscr{M}_\text{FD} :=\ln \left(\cfrac{P}{\lambda-\lambda^{-2}}\right),
\qquad
\zeta:= \lambda^2 + 2\lambda^{-1} -3\thinspace.
\end{equation}

\begin{figure}[H]
    \advance\leftskip-2cm
             \includegraphics[width=20cm, height=7.10cm]{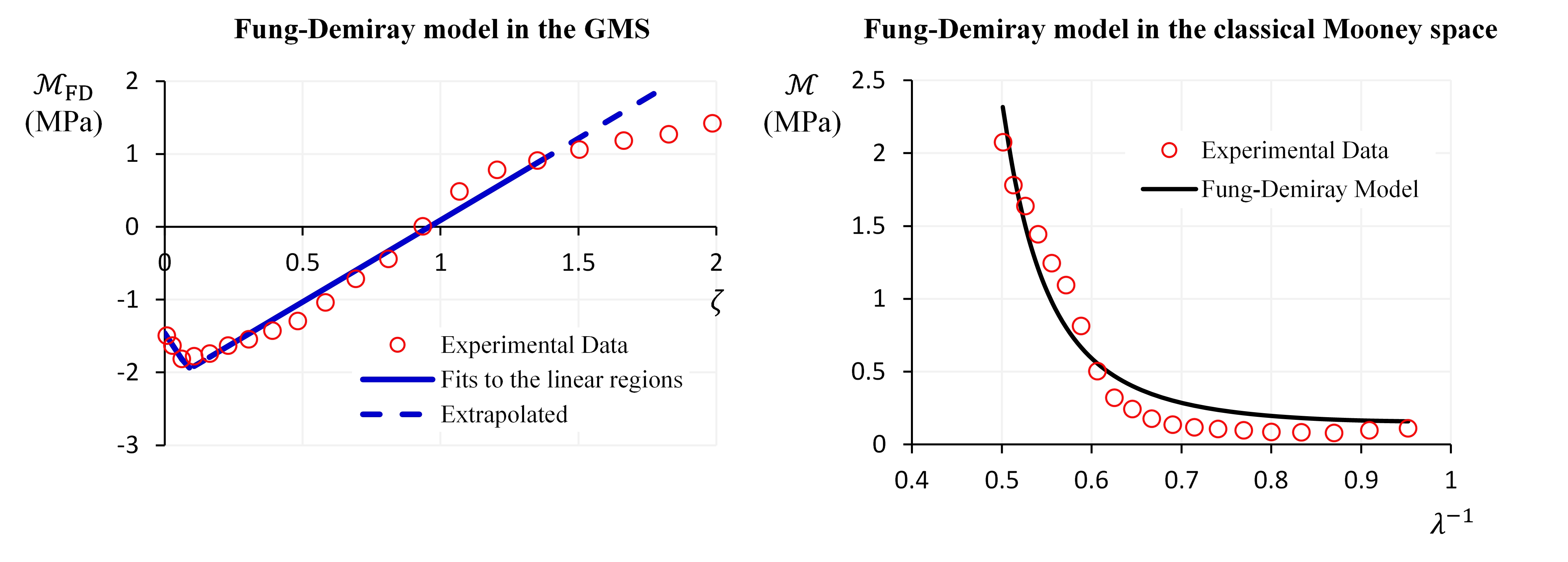} 
     \caption{The transformation of the Fung-Demiray model in the GMS (left panel) and the classical Mooney space (right panel). 
A V-shaped pattern in the experimental data is again observed within the GMS, while the classical Mooney space does not delineate this trend. 
The Fung-Demiray model is able to provide a linear fit to either branches of this V-shaped trend, but clearly not to the whole range of the data. 
The blue lines show the provided best lines of fit to the data in each branch of this V.
Experimental data is from the uniaxial deformation of a healthy porcine aorta.  
}
     \label{Fig4}
\end{figure}

The plots in Figure \ref{Fig4} present the fitting results of this model to the experimental data obtained from healthy porcine aorta under uniaxial deformation, within both the GMS and the classical Mooney spaces (data recorded by Michel Destrade at University College Dublin as part of a separate prior study, also used in the paper by \citeauthor{destrade2009bending} \citeyear{destrade2009bending}).


Similar to the trend observed for the modelling results pertaining to the Gent model, a V-shaped pattern in the experimental data is highlighted again in the GMS. 
As Equation \eqref{Fung2} demonstrates, the Fung-Demiray model can only capture the whole data as a single line. 
However, the GMS allows the identification of the V-shaped pattern in the data, \color{black}and so enables the fitting of the model to either branches separately as a straight line.
The apex of the V is located at $\zeta \simeq 0.06$, corresponding to $\lambda \simeq 1.15$.   
See the blue lines in Figure \ref{Fig4}(a). 
However, note that the gradient of the model line for $\zeta\leq 0.06$  is negative, requiring the stiffening parameter $b$ to be also negative if the Fung-Demiray model is to provide a suitable fit to the data at the smaller range of deformation.
A negative stiffening parameter $b$ \color{black}may not be deemed physically realistic, and therefore caution must be exercised on using the Fung-Demiray strain energy function to model the deformation of isotropic soft tissues. 
Again, note that this behaviour is not captured in the classical Mooney space, see Figure \ref{Fig4}(b).  


\section{Other classes of deformation}
\label{Section-other-deformations}


In the previous sections we presented the concept of the GMS by appealing to the simple tension deformation as a descriptive example. 
However, the application of the GMS is not restricted to uniaxial, nor to homogeneous, deformations. 
As a general rule, when the deformation depends only on one kinematic variable we may find a $\mathscr{M}$ function for linear regression in a similar manner to that devised for the uniaxial deformation case. 

Here, accordingly, we extend the application of the GMS to other classes of deformation: equi-biaxial tension, pure shear, simple shear and torsion. 
The derivations of the $\mathscr{M}$ and $\zeta$ domains for the archetypical strain energy functions considered in this study under the foregoing deformations are similar to those presented in \S2.
For brevity, here we only present the final results, see Table \ref{Table1}.

Similarly, our analysis of the condition number $\kappa$ (not reproduced here) shows that, as in the case of simple extension, the fits obtained in the GMS for these deformations are also \textit{a priori} more robust than those achieved in the Cauchy or engineering spaces, since the values of $\kappa_\mathscr{M}$ for all the foregoing deformation modes are bounded to far lower values than those of $\kappa_T$ and $\kappa_P$.  

\begin{table}[h]
\scriptsize
\begin{tabular}{ c | c | c | c | c | c  }
                   & \textbf{Mooney-Rivlin} & \textbf{Yeoh} & \textbf{Gent-Thomas} & \textbf{Gent} & \textbf{Fung}
                                                         \\ \hline\hline \\[-8pt]
   &  & $C_1(I_1-3)$ &  &  &  
   \\[-3pt]
 $W$ & $C_1(I_1-3)$ & $+ C_2(I_1-3)^2$ & $C_1(I_1-3)$ & $\dfrac{\mu}{2J_m} \ln\left(1-\frac{I_1-3}{J_m}\right)$ & $\dfrac{\mu}{2b}\left[e^{b(I_1-3)}-1\right]$ \\
         & $+ C_2(I_2-3)$ & $+ C_3(I_1-3)^3$ & $+ C_2\ln\left(\frac{I_2}{3}\right)$  &  &  
 \\[-8pt] \\ \hline \\[-8pt]
  \multicolumn{6}{c}{\emph{uniaxial tension}: engineering stress $P$, stretches $\lambda_1=\lambda$, $\lambda_2=\lambda_3=\lambda^{-1/2}$} 
  \\ \hline \\[-8pt]
    fit & $C_1 + C_2 \zeta$ & $C_1 + 2C_2 \zeta + 3C_3\zeta^2$ & $C_1 + C_2 \zeta$ & $\mu + \frac{1}{J_m} \zeta$ & $\ln \mu + b \zeta$\\[-8pt]  
    \\  \\[-8pt]
  $\mathscr{M}$ & $\dfrac{P}{2(\lambda - \lambda^{-2})} $ & $\dfrac{P}{2(\lambda - \lambda^{-2})} $ & $\dfrac{P}{2(\lambda - \lambda^{-2})} $ & $\dfrac{P}{\lambda - \lambda^{-2}} $ & $\ln\left(\dfrac{P}{\lambda - \lambda^{-2}} \right)$ 
  \\ \\[-8pt]
$ \zeta $ &  $ \lambda^{-1} $ &  $\lambda^2 + 2\lambda^{-1} -3$ &  $\dfrac{1}{2\lambda^2 + \lambda^{-1}}$ & $ \cfrac{P(\lambda^2 + 2\lambda^{-1} -3)}{\lambda - \lambda^{-2}}$ & $\lambda^2 + 2\lambda^{-1} -3$
\\[-8pt] \\ \hline \\[-8pt]
 \multicolumn{6}{c}{\emph{equi-biaxial tension}: engineering stress $P$, stretches $\lambda_1= \lambda_2 = \lambda$, $\lambda_3=\lambda^{-2}$} 
 \\ \hline \\[-8pt]
    fit & $C_1 + C_2 \zeta$ & $C_1 + 2C_2 \zeta + 3C_3\zeta^2$ & $C_1 + C_2 \zeta$ & $\mu + \frac{1}{J_m} \zeta$ & $\ln \mu + b \zeta$
    \\[-8pt]    \\  \\[-8pt]
 $\mathscr{M}$ & $\dfrac{P}{2(\lambda - \lambda^{-5})} $ & $\dfrac{P}{2(\lambda - \lambda^{-5})} $ & $\dfrac{P}{2(\lambda - \lambda^{-5})} $ & $\dfrac{P}{\lambda - \lambda^{-5}} $ & $\ln\left(\dfrac{P}{\lambda - \lambda^{-5}} \right)$ 
  \\ \\[-8pt]
$ \zeta $ &  $ \lambda^{2} $ &  $ 2\lambda^2 + \lambda^{-4} -3$ &  $\dfrac{1}{\lambda^2 + 2\lambda^{-4}}$ & $ \cfrac{P(2\lambda^2 + \lambda^{-4} -3)}{\lambda - \lambda^{-5}}$ & $2\lambda^2 + \lambda^{-4} -3$
\\[-8pt] \\ \hline \\[-8pt]
 \multicolumn{6}{c}{\emph{pure shear}: engineering stress $P$, stretches $\lambda_1= \lambda$, $\lambda_2 = 1$, $\lambda_3=\lambda^{-1}$}
  \\ \hline \\[-8pt]
    fit & $C_1 + C_2 $ & $C_1 + 2C_2 \zeta + 3C_3\zeta^2$ & $C_1 + C_2 \zeta$ & $\mu + \frac{1}{J_m} \zeta$ & $\ln \mu + b \zeta$
    \\[-8pt]  \\  \\[-8pt]
$\mathscr{M}$ & $\dfrac{P}{2(\lambda - \lambda^{-3})} $ & $\dfrac{P}{2(\lambda - \lambda^{-3})} $ & $\dfrac{P}{2(\lambda - \lambda^{-3})} $ & $\dfrac{P}{\lambda - \lambda^{-3}} $ & $\ln\left(\dfrac{P}{\lambda - \lambda^{-3}} \right)$ 
  \\ \\[-8pt]
$ \zeta $ &  $  $ &  $ \lambda^2 + \lambda^{-2} - 2$ &  $\dfrac{1}{\lambda^2 + \lambda^{-2} + 1}$ & $ \cfrac{P(\lambda^2 + \lambda^{-2} - 2)}{\lambda - \lambda^{-3}} $ & $\lambda^2 + \lambda^{-2} - 2$
\\[-8pt] \\ \hline \\[-8pt]
 \multicolumn{6}{c}{\emph{simple shear}: shear stress $T$, amount of shear  $\gamma$}
  \\ \hline \\[-8pt]
    fit & $C_1 + C_2 $ & $C_1 + 2C_2 \zeta + 3C_3\zeta^2$ & $C_1 + C_2 \zeta$ & $\mu + \frac{1}{J_m} \zeta$ & $\ln \mu + b \zeta$
    \\[-8pt]  \\  \\[-8pt]
$\mathscr{M}$ & $\dfrac{T}{2\gamma} $ & $\dfrac{T}{2\gamma} $ & $\dfrac{T}{2\gamma} $ & $\dfrac{T}{\gamma}$ & $\ln\left(\dfrac{T}{\gamma} \right)$ 
  \\ \\[-8pt]
$ \zeta $ &  $  $ &  $ \gamma^2$ &  $\dfrac{1}{\gamma^2+3}$ & $T\gamma$ & $\gamma^2$
\\[-8pt] \\ \hline \\[-8pt]
 \multicolumn{6}{c}{\emph{simple torsion}: axial load $\mathscr{N}$, twist per unit undeformed length $\phi$, undeformed radius $R_0$}
  \\ \hline \\[-8pt]
    fit & $-\cfrac{\pi}{2}\left(C_1 +2 C_2\right) $ & $-\cfrac{\pi}{12}\left(6C_1 +8 C_2\zeta+9C_3\zeta^2\right)$ & $-\cfrac{\pi}{2}\left(C_1 +4 C_2\zeta_{\text{GT}}\right)$ & $\mu + \frac{1}{J_m} \zeta_\text{G}$ & $\mu + b \zeta_{\text{FD}}$
    \\[-8pt]  \\  \\[-8pt]
$\mathscr{M}$ & $ \cfrac{\mathscr{N}}{R_o^4\thinspace\phi^2} $ & $\cfrac{\mathscr{N}}{R_o^4\thinspace\phi^2}$ & $ \cfrac{\mathscr{N}}{R_o^4\thinspace\phi^2} $ & $\cfrac{\mathscr{N}}{R_o^4\thinspace\phi^2}$ & $\cfrac{\mathscr{N}}{R_o^4\thinspace\phi^2}$ 
  \\ \\[-6pt]
$ \zeta $ &  $ R_o^2\thinspace\phi^2 $ & $R_o^2\thinspace\phi^2$ &  $ \zeta_{\text{GT}} $ & $\zeta_{\text{G}}$ & $\zeta_{\text{FD}}$
  \\[2pt] \hline 
\end{tabular}
\caption{GMS for the archetypical strain energy functions $W$ considered in this study: fitting objectives and the transformed domain variables $\mathscr{M}$ and $\zeta$.\\}
\label{Table1}
\end{table}

\noindent In Table \ref{Table1}, $\zeta_{\text{GT}}$, $\zeta_{\text{G}}$ and $\zeta_{\text{FD}}$ for simple torsion  are defined as:
\begin{align} 
&  \zeta_{\text{GT}}=\cfrac{1}{R_o^2\phi^2}\left[1-\cfrac{3}{R_o^2\phi^2}\text{ln}\left(\cfrac{R_o^2\phi^2+3}{3}\right)\right], \label{eq30N}
\\[12pt]   
&  \zeta_{\text{G}}=\cfrac{\mathscr{N}}{R_o^4\phi^2}\thinspace J_m \left\{1+\cfrac{2R_o^4\phi^4}{\pi\left[\text{ln}\left(\cfrac{R_o^2\phi^2-J_m}{J_m}\right)+\cfrac{R_o^2\phi^2}{J_m}\right]}\right\}, \label{eq31N}
 \\[12pt]       
  &  \zeta_{\text{FD}}=\cfrac{\mathscr{N}}{R_o^4\phi^2}\left\{1+\cfrac{2b^2R_o^4\phi^4}{\pi\text{exp}(-1)\left[\left(bR_o^2\phi^2-1\right)\text{exp}(bR_o^2\phi^2)+1\right]}\right\}. \label{eq32N}
\end{align}


\section{Concluding remarks}
\label{conclusion}


Our aim in this paper was to put forward a systematic view of the mathematics and mechanics at play in the Mooney plot. 
Our analysis highlights that in the arena of modelling the finite deformation of rubber-like materials, there is not a single Mooney plot, but  many Mooney plots, depending on the model chosen to fit the data and on the mode of deformation. These Mooney plots, however, may all be constructed under a canonical concept which we have coined in this paper as the \emph{generalised Mooney space} (GMS). 

To the best of our knowledge, a clear and explicit rationalisation of the underlying reason(s) that prompted \cite{RS} to introduce and use the \textit{classical} Mooney plot has not been articulated in the literature. 
While the Mooney plot was not used in the seminal work of \cite{Mooney} itself, we can nonetheless argue that it was a helpful tool from a computational point of view to facilitate a linear regression problem with much simpler calculations.
Since its inception, the Mooney plot has been used by several researchers to provide an \textit{ad primum aspectum} of the goodness of a fit, which often remains hidden in the Cauchy or engineering spaces. 
A typical example is provided by the neo-Hookean model, which in the engineering space seems to provide a satisfactory fit to the finite but moderate deformation ranges of rubbers, only then to demonstrate significant shortcomings in the Mooney plot, coupled with a poor performance in respect of minimising the relative errors (see, e.g., \citeauthor{MethodicalFit} \citeyear{MethodicalFit}). 

The usefulness of the GMS reveals itself first as an improvement on the fitting process. 
Although computational power has increased dramatically since the early days of research on rubber, nonlinear curve fitting exercises are fraught with potential pitfalls, as uniqueness of an optimal set of parameters is not guaranteed \citep{OgdenFitting, MethodicalFit}. 
In the examples presented here, we saw that the GMS turns a nonlinear fitting exercise into a linear one for a variety of models including the Gent and the Fung-Demiray models.
Another clear advantage of the GMS lies in its ability to identify the limitations of a given model. 
This feature arises as a result of the bounded condition numbers $\kappa$ pertaining to the functional forms of $\mathscr{M}$ for the models transformed into the GMS. 
As exemplified by the archetypical Mooney-Rivlin model in \S2, for many models, if not all, one can notice the presence of kinematical factors in the stress-strain relationships within the Cauchy or engineering spaces that  cause the condition number $\kappa$ to approach infinity at ranges close to the unstrained state. 
This phenomenon hides the descriptive ability of the model in the small to moderate ranges of the deformation within the Cauchy or engineering spaces. 
The GMS, by contrast, solves this problem by providing $\mathscr{M}$ functions that have a bounded condition number $\kappa_\mathscr{M}$ over the whole range of deformation. 
\color{black}In other words, the GMS appears to `{\emph{clean}}' the kinematical sensitivities present within the classical Cauchy or engineering spaces; an outlook similar to the approach of \cite{Criscione} using a new domain of invariant space. 
\color{black}As a result, the performance of a model and its limitations can be observed with a better magnification, and with more robustness. 

In view of the foregoing, and of the analyses and results presented in this work, we conclude that the GMS allows for the possibility of gaining meticulous insights into the performance of different models in ways that are not easily accessible in the current classical spaces. For this reason we believe that the GMPs are a new class of tools that enable a clearer and more quantitative assessment of the performance of many models of the nonlinear elasticity theory. Given that the variety of the proposed models continually increases in the literature, this new tool may provide a fundamental assistance to understand the focal points in solving the \textit{Hauptproblem}.


\section*{Acknowledgments}


The research of MD was supported by a 111 Project for International Collaboration (Chinese Government, PR China) No.~B21034 and by a grant from the Seagull Program (Zhejiang Province, PR China). GS is supported by the Istituto Nazionale di Alta Matematica (INdAM, Italy), Fondi di Ricerca di Base UNIPG (Italy), the MIUR-PRIN project 2017KL4EF3 and Istituto Nazionale di Fisica Nucleare through its IS ‘Mathematical Methods in Non-Linear Physics’
(Italy).



\newpage 

\bibliography{RefGMSPaperRevised} 


\end{document}